\documentclass[english,aps,pra,reprint,noshowpacs]{revtex4-1}   
\usepackage[T1]{fontenc}	% should generally be included for better accented-word behavior

\usepackage[latin9]{inputenc}	% should generally be included for better accent behavior

\usepackage{geometry}		% for controlling page margins

\usepackage{graphicx}

\usepackage[above,below]{placeins}	% allows use of \FloatBar�rier command to force section barriers
\usepackage{times}
\usepackage{amsmath}
\usepackage{multirow}

\usepackage[textwidth=4.5cm]{todonotes}

\usepackage{multirow}

\usepackage{lipsum}
\usepackage{titlesec}
\titlespacing\section{0pt}{10pt plus 4pt minus 2pt}{10pt plus 2pt minus 2pt}
\titlespacing\subsection{0pt}{10pt plus 4pt minus 2pt}{10pt plus 2pt minus 2pt}
\begin{document}

\title{Examining Student Participation in Two-Phase Collaborative Exams through Video Analysis}
\author{Joss Ives$^1$, Matias de Jong Van Lier$^1$, Nutifafa Kwaku Sumah$^1$, and Jared B. Stang$^1$}
\affiliation{$^1$Dept. of Physics \& Astronomy, University of British Columbia, 6224 Agricultural Road, Vancouver, BC, V6T 1Z1}

%\keywords{}

\begin{abstract}
In this study we coded, for individual student participation on each question, the video of twenty-seven groups interacting in the group phase of a variety of two-phase exams. We found that maximum group participation occurred on questions where at least one person in the group had answered that question incorrectly during the solo phase of the exam. We also observed that those students that were correct on a question during the solo phase have higher participation than those that were incorrect. Finally we observed that, from a participation standpoint, the higher-scoring (lower-scoring) students seem to benefit the most (least) from heterogeneous groups, while homogeneous groups do not seem to favor students of any particular performance level. 

\end{abstract}

\maketitle

%============================================================
%============================================================
\section{INTRODUCTION}
%============================================================
%============================================================

A two-phase (or two-stage) exam is one in which students first take the exam individually (the \textit{solo stage}), hand in their solo exams, and then immediately re-take the exam in small collaborative groups (the \textit{group stage}) of three or four. This assessment strategy builds on the collaborative small-group instructional strategies used in many interactive engagement classrooms \cite{FreemanPNAS2014}, such as the use of clickers and worksheets.

Previous studies have found that two-phase exams improve content retention \cite{Gilley2014}, improve short-term performance on matched near-transfer questions \cite{Ives2014}, and have many positive affective benefits \cite{Rieger2014, Leight2012}. However, little is known about the interactions that students are having during the group stage of these exams. Beatty \cite{Beatty2015} made use of student response patterns to infer the quality of the group interactions and others \cite{James2011} have used audio to learn about the discussions that take place during in-class small-group interactions, an instructional strategy which can be viewed as a low-stakes implementation of two-phase exams. 

An important measure of quality of group interaction is a high level of participation from each student in the group. In this study we coded, for individual student participation, the video of twenty-seven groups interacting in a variety of formal assessment situations. Quantifying student participation allows us to examine the factors that influence the experience of students during the two-phase exam, which in turn can inform two-phase exam practices. This research study comprises a first step towards understanding how both test design and group structure impact participation during the group phase of the exam. To this end, we examine several specific questions, such as ``How do interactions during the group phase relate to the difficulty of the question for that group?'' and ``How does a student's performance on the solo stage correlate with their participation during the group stage?''

%============================================================
%============================================================
\section{METHODS - CONTEXT}
%============================================================
%============================================================

%------------------------------------------------------------
%\subsection{Context}
%------------------------------------------------------------

Participants were recruited from two different calculus-based introductory Physics courses at the University of British Columbia. 

Course A, an introductory mechanics course, had an enrollment of 83, all of whom were international students. In each of six every-other-week tests during the term, the group stage for two to three groups of study participants was video-recorded in the same examination room as all students taking the test, for a total of thirteen recorded groups. The solo stage of these tests consisted of an hour-long test with multiple-choice, other short-answer questions, as well as one or two longer problems. Questions were converted to five-answer multiple-choice questions for the group phase of the test. 

Course B, an introductory fluids, waves and energy course, had an enrollment of 715, which consisted of a mix of international and domestic students. This course used two-phase exams for both midterms and the final exam, with three to seven groups being video-recorded on a given exam, for a total of fourteen recorded groups. In this multi-section course, each exam was invigilated in multiple different rooms at the same time. In order for the researchers to be able to set up the video recording equipment for multiple groups at the same time, the participants in course B were invited to take their exams in a room consisting only of study participants. The exam format in this course was similar to that of course A, with a slightly heavier emphasis on longer problems. Same as course A, all questions were converted to five-answer multiple choice questions for the group phase of the exam.

\section{METHODS - CODING OF VIDEO}
%------------------------------------------------------------
%\subsection{Coding of Video}
%------------------------------------------------------------

For each group each question was coded using two different methods. First, the participation level of each member of the group was scored. Second, an interaction category was assigned to describe the type of interaction that the group had. In addition to describing the interaction categories and the descriptors used to assign participation scores, development of the coding schemes is also described in this section.

Video was coded for participation on a 0-3 point scale for each member of a group for each question. Table \ref{table-participation-rubric} shows a summary of the descriptions of an individual student's behaviors related to each possible score on the participation rubric. This summary table is intended to provide the reader with basic interpretation of the participation score levels. The full rubric used by those that coded the video had much more detail for the descriptors at each level.

% For a two-column table, use \begin{table*} instead (with matching \end{table*}.
\begin{table}[tp]
\caption{Summary of the participation rubric descriptors.}
\label{table-participation-rubric}
\begin{ruledtabular}
\begin{tabular}{cl}
Score & Description of student's participatory behavior \\
 \hline
3 & Asks questions or provides answers with explanations \\
2 & States their answer or assists in explanation \\
1 & Visibly engaged but silent or intermittently engaged \\
0 & No interactions with the group
\end{tabular}
\end{ruledtabular}
\end{table}
% Formatting hack if needed--FloatBarrier forces floats to show here, before next section
%\FloatBarrier	

We noticed in the development of the participation rubric that the group interactions fell into three distinct interaction categories, which we have named \textit{A1}, \textit{AA} and \textit{MA} and are described in Table \ref{table-interaction-categories}. From the perspective of wanting to maximize the number of people with a high level of participation in the group conversations, the \textit{MA} interaction type is considered to be the most desirable.

\begin{table}[bp]
\caption{Summary of the interaction categories.}
\label{table-interaction-categories}
\begin{ruledtabular}
\begin{tabular}{cl}
Code & Description of group interaction \\
 \hline
\textit{A1} & \multirow{2}{0.9\linewidth}{One person dominates the group discussion and no other members of the group ask any questions of significance.} \\
& \\
\textit{AA} & \multirow{3}{0.9\linewidth}{Multiple members of the group state their answer from the solo phase and the group chooses their answer without any significant discussion.} \\
& \\
& \\
\textit{MA} & \multirow{3}{0.9\linewidth}{A variety of interaction types where multiple group members are interacting using a combination of asking questions, offering different viewpoints or explaining their reasoning.} \\
& \\
& \\
\end{tabular}
\end{ruledtabular}
\end{table}

Development of the rubrics and coding of the video was done by three of the authors (NKS, MVL and JI). Initial development of the rubrics was done by NKS, with later refinement by MVL. JI participated in the rubric development stage and acted as a third coder for inter-rater reliability checks. Three sets of inter-rater reliability checks were performed with the fully-developed rubrics. Between NKS and JI, 29 questions were coded for a total of 116 participation scores assigned. Before discussing disagreements, the inter-rater reliability for interaction categories was 93.1\% and for participation scores was 88.8\%, with all disagreements resolved after discussion. Between MVL and JI, 19 questions were coded for a total of 76 participation scores assigned. Before discussing disagreements, the inter-rater reliability for interaction categories was 84.2\% and for participation scores was 85.5\%, with all disagreements resolved after discussion. Although they did not work on the project at the same time, MVL performed inter-rater reliability checks between his own coding and previous coding by NKS. Between those two coders, 30 questions were coded for a total of 120 participation scores assigned. The inter-rater reliability for interaction categories was 83.3\% and for participation scores was 81.7\%.

%============================================================
%============================================================
\section{RESULTS AND DISCUSSION}
%============================================================
%============================================================ 

The participation scores presented in this manuscript are ordinal, meaning that they are ranked (i.e., 3 represents higher participation than 2), but that the spacing between the values may not be the same across all levels. Despite this fact,  we use standard error of the mean (SEM) throughout the figures in this manuscript to represent a coarse depiction of the confidence in the mean of participation scores in order to be able use our data representations to make rough visual comparisons. These SEMs cannot be interpreted to imply proper 68\% confidence intervals. 

We will be conservative throughout the manuscript, commenting only on trends or differences that are obvious to the eye (i.e., large effect sizes) and do not require statistical tests to demonstrate differences. In order to earn the confidence of the reader regarding this approach, we run a Mann-Whitney U test for select comparisons in the manuscript. See, for example, Figures \ref{fig-correct_vs_incorrect} and \ref{fig-participation_vs_nCorrect}.

%({\color{red}\textit{Perhaps show histogram?}})

%{\color{red}\textit{Also need to add a caveat that the participation scores are ordinal. We could run some tests to show if the data are normally distributed. When we do statistical comparisons we will run Mann-Whitney U, which is fine for ordinal data, but we use standard deviation and standard error for a lot of uncertainty representations, which is not something you should be able to do with ordinal data. Cliff's delta is used for effect size}} %https://en.wikipedia.org/wiki/Effect_size#Effect_size_for_ordinal_data

%------------------------------------------------------------
\subsection{Frequency of interaction categories}
%------------------------------------------------------------
% Joss: Needs a better section title

How frequently do we see each category of group interaction and do these frequencies depend on the percentage of students correct on that question? 

\begin{figure}[bp]
\includegraphics[width=1.0\linewidth]{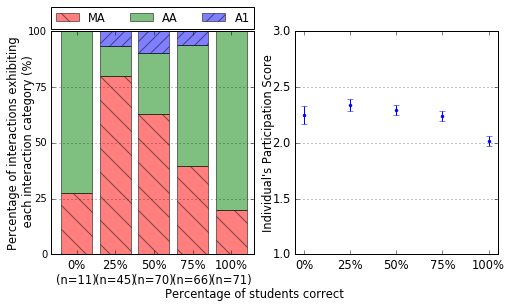}
%{images/Pa_vs_per_correct_and_int_type_for_perc_correct_JOINED_V2}
\caption{Percentage of interactions exhibiting each interaction category broken down by the fraction of students in the group answering the question correctly on the solo phase of the exam (left) and participation scores for fraction of students correct (right). Groups of three were not included in these plots due to the small number of questions coded (N=53) for that group size. In the left panel, the total number of interactions coded for each category of percentage correct is included under the column label. Our participation score range runs from 0 to 3, but we display only from 1 to 3 throughout the paper in order to better visualize trends in the data.}
\label{fig-interaction_breakdown}
\end{figure}

Figure \ref{fig-interaction_breakdown} (left panel) shows that the \textit{A1} interaction is infrequent (occurring less than 6\% of the time overall) and the \textit{AA} and \textit{MA} occur with similar frequency, each close to half of the time. However, looking at the breakdown of interaction categories as a function of the percentage of students correct on that question of their solo phase exam reveals that this breakdown shows an extremely strong dependence on percentage of students correct. Groups with at least one correct student, but no more than half appear to be the most productive from the perspective of maximizing the group's likelihood of engaging in an \textit{MA} interaction. The right panel of this figure shows that participation scores are noticeably lower for groups where everybody was correct on that question.

%------------------------------------------------------------
\subsection{Impact of an Individual Student Being Correct}
%------------------------------------------------------------

% * <jossives@gmail.com> 2016-06-28T19:46:44.900Z:
%
% In order to establish that MA is the most desirable type of interaction, from a participation point of view, do we need to use statistics to show that MA is way above and to establish the correct participates more at a statistically significant effect size? How do we combine the effect of correct vs incorrect with th`e interaction type. Looking at this section's figure, the take-home message is that correct vs incorrect is a second-order effect behind the interaction category
%
% ^.

How does an individual student's participation in the group discussion for a given question depend on their being correct in the solo phase? Does this also depend on the fraction of students correct in the group?

\begin{figure}[tp]
\includegraphics[width=1.0\linewidth]{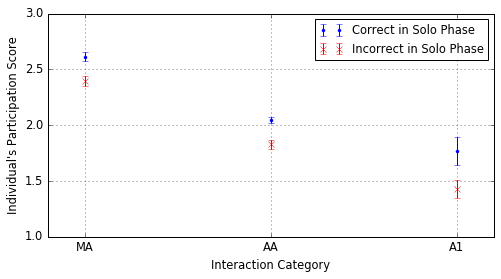}
\caption{Comparison of the participation scores for individual students when they were correct or incorrect on that question in the solo phase, broken down by interaction category. Uncertainties are SEM. As an example of quantifying differences that appear obvious to the eye, the mean participation scores in the \textit{MA} interaction category were 2.61 (correct) and 2.39 (incorrect); the distributions in the two groups differed significantly (Mann-Whitney \textit{U} = 50456, $n_{\textrm{correct}} = 290$, $n_{\textrm{incorrect}}=297$, $p=2.4\times 10^{-5}$) with an effect size of $d$ = .17, calculated using Cliff's delta. }
\label{fig-correct_vs_incorrect}
\end{figure}

Figure \ref{fig-correct_vs_incorrect} shows that, across all three interaction categories, individual students participate more when they have answered a question correctly in the solo phase versus when they have not. We also see from this figure that the participation level in \textit{MA} interactions, independent of correctness, is higher than in the other two interactions types.
%Figure \ref{fig-correct_vs_incorrect} shows that, across all three interaction categories, individual students participate more when they have answered a question correctly in the solo phase versus when they have not. Additionally, we see from this figure that the participation level in \textit{MA} interactions, whether the student was correct or incorrect, is higher than in the other two interactions types.

\begin{figure}[bp]
\includegraphics[width=1.0\linewidth]{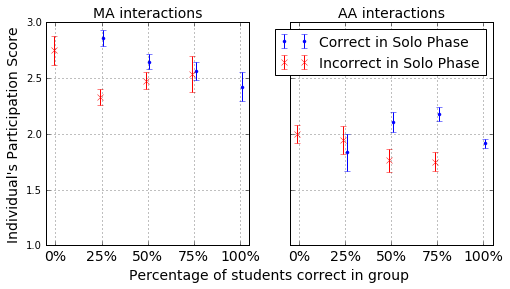}
\caption{Participation scores as a function of percentage of students correct in a group. Comparisons are made between those that have answered a question correctly in the solo phase versus when they have not, looking at \textit{MA} interactions (left panel) and \textit{AA} interaction (right panel). These results include only students from groups of 4. Uncertainties are SEM. As an example of quantifying differences that appear obvious to the eye, the means for \textit{AA} interactions for 50\% of students correct were 2.11 for correct and 1.76 for incorrect; the distributions in the two groups differed significantly (Mann-Whitney \textit{U} = 925, $n_{\textrm{correct}} = n_{\textrm{incorrect}} = 38$, $p=.015$) with an effect size of $d$ = .28, calculated using Cliff's delta.
}
\label{fig-participation_vs_nCorrect}
\end{figure}

We can further subdivide these data and look at how a student's participation, when they were correct or incorrect, varies with respect to the fraction of the people in the group that answered that question correctly. When looking at cases where both correct and incorrect people are present in the group (percentage correct from 25\% to 75\% in Figure \ref{fig-participation_vs_nCorrect}), we see that the trends are the opposite for \textit{MA} and \textit{AA} interactions. In \textit{MA} interactions average participation is even between correct and incorrect group members for groups with 75\% of the students correct. However, as the percentage of students correct in the group drops, we see that the participation of the correct students becomes much larger than the participation of incorrect students. 

In \textit{AA} interactions, average participation is consistent between correct and incorrect group members for groups with 25\% of the students correct. However, as the percentage of students correct in the group increases, we see that the participation of the correct students becomes much larger than the participation of incorrect students. 

Overall we see that across all interaction categories correct students participate more than incorrect students. Exceptions are \textit{MA} interactions with a majority of students correct and \textit{AA} interactions with a minority of students correct.

%({\color{red}\textit{Joss: Almost certainly will need an effect size and significance for this}}). {\color{red}\textit{Describe the AA graph and add some narrative.}}

%------------------------------------------------------------
\subsection{Heterogeneity}
%------------------------------------------------------------

How does an individual student's participation in the group depend on the heterogeneity of test performances within the group? Does this impact high and low-scoring students differently?

In order to look at the effect of heterogeneity, we used the standard deviation of the solo phase grades within the group. This resulted in the twenty-seven groups being placed in three levels of heterogeneity, with nine groups in each: Low (standard deviations between 3.7\% and 9.0\%, inclusive), High (standard deviations between 18.6\% and 30.5\%, inclusive), and Medium (those in between Low and High).

\begin{figure}[tp]
\includegraphics[width=1.0\linewidth]{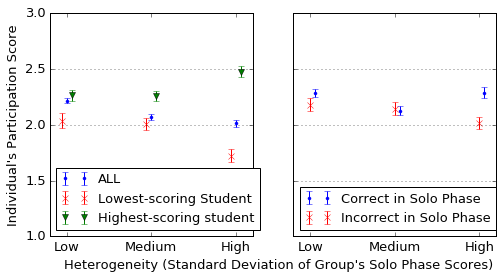}
\caption{
Participation scores as a function of group heterogeneity, as measured by the standard deviation of the group's solo phase scores. The scores in the right panel are a subset of those from the left panel because questions from the solo phase that involved partial marks (e.g., longer computational problems) were not included in the correct/incorrect part of the analysis. All uncertainties are SEM.
%When you look at highest scoring student in the group, they participate more as the standard deviation of the group increases. When you look at the lowest scoring student in the group, they participate less as the standard deviation of the group increases
}
\label{fig-participation_vs_SD}
\end{figure}

When we look at the relationship between participation and group heterogeneity (see Figure \ref{fig-participation_vs_SD}, left panel), we see that average participation score of the group tends to decrease with increased heterogeneity. We also find this trend holds for the lowest scoring student in the group, as measured by their score in the solo phase. However, when looking at the highest scoring student in the group, we see that their participation increases with increasing heterogeneity. We also found that these same trends held if we looked at only questions with \textit{MA} interactions or only questions with \textit{AA} questions. A plausible explanation for these trends is that the highest scoring member of the group takes on more of a leadership role when the performance levels in the group are more diverse, but in the more homogeneous groups, the leadership role is shared by many group members. 

Consistent with the explanation that the higher scoring students take on a larger leadership role in the groups with higher levels of heterogeneity, the right panel of Figure \ref{fig-participation_vs_SD} shows that, in groups of Low and Medium heterogeneity, there is no significant difference between students that answered a question correct or incorrect, but in the groups of High heterogeneity, the students with the correct answer participate more.

In general we see small decreases in participation as the heterogeneity of the group increases. However, we see that higher scoring students (those in upper quartiles or those that answered a question correctly) have increased participation with increasing heterogeneity, suggesting that they take on a larger leadership role in the more heterogeneous groups.

%------------------------------------------------------------
\subsection{Overall Solo Phase Performance}
%------------------------------------------------------------

How does a student's participation in the group discussion correlate with their performance from the solo phase, both overall and relative to the other students in their group?

\begin{figure}[b]
\includegraphics[width=1.0\linewidth]{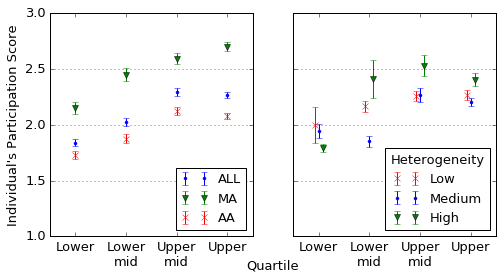}
\caption{
Participation scores as a function of student performance on the solo phase of the exam, as grouped by quartile on that exam. The right panel shows these data grouped by group heterogeneity, as measured by the standard deviation of the group's solo phase scores. All uncertainties are SEM.
}
\label{fig-participation_vs_quartile}
\end{figure}

Since the data collected in this study come from a variety of different tests across two courses, we have chosen to use quartiles to communicate how an individual student performs relative to their peers on a given test. The left panel of Figure \ref{fig-participation_vs_quartile} shows a strong positive correlation between participation and test performance relative to one's peers, independent of interaction category. Although not included in this manuscript, it was found that a plot of participation versus rank in the group (based on solo phase test score) showed very similar trends. 
The right panel of Figure \ref{fig-participation_vs_quartile} shows that homogeneous groups appear to have the flattest participation profiles across student performance levels at the test or individual question levels.

%============================================================
%============================================================
\section{CONCLUSIONS}
%============================================================
%============================================================ 

We coded video recordings of twenty-seven groups for individual student participation while taking a group exam.
We found that a diversity of opinions in the group (as measured by percentage of students correct in the group) is the best way to maximize overall group participation. This suggests that group exam questions should be fairly difficult, but one has to be careful about making them too difficult since groups with none of the students correct in the group rarely participate in interactions of the type that maximize overall participation. 
The results are ambiguous with respect to homogeneous or heterogeneous groups being most favorable. As compared to homogeneous groups, heterogeneous groups see higher participation from those with higher test scores and those that were correct on a given question. 
%However, homogeneous groups appear to have the flattest participation profiles across student performance levels at the test or individual question levels.
However, homogeneous groups do not seem to favor students of any particular performance level.

We have only scratched the surface on the possibilities of this data set. Coding is currently underway to characterize the types of interactions beyond a single participation score, such as tracking who has asked a question or provided an explanation. We also plan to look at gender effects, as well as the effect of difficulty level and different question types.

%============================================================
%============================================================ 

\end{document}